\documentclass{article}

\usepackage[english]{babel}

\usepackage[letterpaper,top=2cm,bottom=2cm,left=3cm,right=3cm,marginparwidth=1.75cm]{geometry}

\usepackage{amsmath,amssymb}
\usepackage{graphicx}
\usepackage[colorlinks=true, allcolors=blue]{hyperref}
\usepackage{authblk}   

\title{A unified framework for hot accretion flows with finite angular momentum: from Bondi-like to disc-like regimes
}
\author[1,2,3]{Cheng-Liang Jiao\thanks{Email: jiaocl@ynao.ac.cn}}
\author[1,4,3]{Liying Zhu}
\author[1,3]{Er-gang Zhao}
\author[1,3]{Xiang-dong Shi}

\affil[1]{Yunnan Observatories, Chinese Academy of Sciences, Kunming 650216, China}
\affil[2]{Center for Astronomical Mega-Science, Chinese Academy of Sciences, Beijing 100012, China}
\affil[3]{Key Laboratory for the Structure and Evolution of Celestial Objects, Chinese Academy of Sciences, Kunming 650216, China}
\affil[4]{University of Chinese Academy of Sciences, Beijing 101408, China}

\begin{document}
\maketitle

\begin{abstract}
Observations of X-ray luminous elliptical galaxies suggest that the accretion rate onto the central supermassive black hole can reach a substantial fraction of the Bondi rate.
However, classical accretion theory applicable to such hot accretion flows treats spherically symmetric Bondi accretion and disc-like advection-dominated accretion flows (ADAFs) as two distinct limiting cases, lacking a unified framework for flows with finite angular momentum.
In this work, we develop such a framework that continuously connects these two regimes.
Our model naturally recovers the Bondi solution in the limit of vanishing angular momentum and approaches the properties of classical ADAFs at high angular momentum, while providing a physically well-defined description of the intermediate regime where neither limiting case is strictly applicable.
We further demonstrate that the accretion rate is jointly regulated by the angular momentum of the ambient gas and the gas viscosity.
For sufficiently large but physically reasonable viscosity, the accretion rate can remain at a significant fraction of the Bondi rate even in the presence of substantial gas rotation.
These results offer a natural explanation for how such accretion rates can be sustained despite finite angular momentum in realistic galactic environments.
\end{abstract}

\section{Introduction}

In the centres of X-ray luminous elliptical galaxies, \emph{Chandra} observations have revealed a tight correlation between the Bondi accretion rate and the mechanical power of relativistic jets inferred from X-ray cavities in the surrounding hot gas.
The Bondi accretion rate ($\dot{M}_\mathrm{B}$), originally derived for spherically symmetric accretion onto a point mass \cite{Bondi1952}, is estimated from observationally constrained black hole masses and the gas density and temperature profiles in the vicinity of the supermassive black holes (SMBHs).
For a sample of nine nearby systems, this relation follows a power-law form,
with the jet power being approximately
$(1.3$--$3.7)\times 10^{-2} \dot{M}_\mathrm{B} c^2$,
and a characteristic value of $\sim 2.2\times 10^{-2} \dot{M}_\mathrm{B} c^2$
\cite{Allen2006}.
Subsequent studies have extended this finding to larger samples of early-type galaxies and brightest cluster galaxies, confirming the correlation over broader dynamic ranges and highlighting its central role in AGN feedback processes that regulate gas cooling and star formation \cite{Fujita2016, Pls2022, Harrison2024}. 
Adopting a jet production efficiency at the level of a few per cent, the observed jet mechanical powers imply a global, order-of-magnitude constraint that the net mass accretion rates onto SMBHs must remain at least a substantial fraction of the Bondi accretion rates.\footnote{
Observational estimates of the jet production efficiency in X-ray luminous radio galaxies indicate values at the level of a few per cent, with a broad distribution across the population \cite{Ich2023}.
Magnetohydrodynamic simulations further indicate that, under typical conditions, the jet production efficiency remains at the level of a few per cent or lower \cite{Mck2009, Pun2011, Ava2016, Mor2018, Mas2023}.
Under special circumstances, such as magnetically arrested discs associated with extreme black hole spins, the efficiency may be substantially enhanced, allowing for lower accretion rates and stronger outflows \cite{Nem2015}.
}
However, in classical accretion theory, the presence of angular momentum tends to suppress accretion relative to the Bondi rate.
In most existing models, even a modest amount of rotation introduces a centrifugal barrier that significantly reduces the mass inflow,
posing a major challenge for reconciling predicted accretion rates with the observed jet powers
\cite{Proga2003, Kru2005, Cuadra2006, Ino2010}.

To address this discrepancy, Narayan and Fabian \cite{NF11} (hereafter NF11) proposed that a slowly rotating, viscous accretion flow forms at the centres of such systems by accreting directly from the hot interstellar medium (ISM). 
They constructed a global, spherically symmetric advection-dominated accretion flow (ADAF) model and demonstrated that both the flow structure and the mass accretion rate depend sensitively on the angular momentum of the ambient gas. 
In particular, they introduced the following two dimensionless parameters,
\begin{equation}\label{L}
\mathcal{R} \equiv \frac{\Omega_\mathrm{out}}{\Omega_\mathrm{K}(r_\mathrm{B})},
\quad 
\mathcal{L} \equiv \frac{l_\mathrm{out}}{l_\mathrm{ms}}
= \frac{r_\mathrm{B}^2 \Omega_\mathrm{out}}{l_\mathrm{ms}},
\end{equation}
where $\Omega_\mathrm{out}$ is the angular velocity of the ambient gas, $r_\mathrm{B}$ is the Bondi radius, and $\Omega_\mathrm{K}(r_\mathrm{B})$ is the Keplerian angular velocity evaluated at $r_\mathrm{B}$. 
Here $l_\mathrm{ms} = \sqrt{27/8}  r_\mathrm{g} c$ denotes the Keplerian specific angular momentum at the marginally stable orbit ($r_\mathrm{ms} = 3 r_\mathrm{g}$; $r_\mathrm{g} = 2GM/c^2$ being the Schwarzschild radius), computed using the pseudo-Newtonian potential \cite{PW1980}. 
For reference, $\mathcal{L} = 1$ corresponds to $\mathcal{R} = 0.0037$, i.e., the gas rotates at only $\sim 0.4\%$ of the Keplerian value.

NF11 found that when $\mathcal{L} \lesssim 1$, the flow becomes Bondi-like: both the flow structure, as characterised by the sonic radius $r_\mathrm{c}$, and the accretion rate converge to their Bondi counterparts. 
As $\mathcal{L}$ increases, the solution undergoes a sharp transition to a disc-like mode, with the sonic radius moving close to $r_\mathrm{ms}$ and the accretion rate dropping significantly. 
This transition occurs at $1.5 \lesssim \mathcal{L} \lesssim 2$. 
Their numerical results further showed that for reasonable values of the viscosity parameter ($\alpha \sim 0.1$--$0.4$  \cite{King2007}) and sufficiently slow rotation ($\mathcal{R} \lesssim 0.3$), the accretion rate can remain relatively high, $\dot{M}_{\mathrm{ADAF}} \sim (0.3$--$1) \dot{M}_{\mathrm{B}}$. 
In this regime, the observed jet power can be explained with an energy efficiency of order $\sim 5\%$, which appears physically plausible.

Despite these successes, the NF11 model assumes spherical symmetry and is therefore strictly applicable only in the Bondi-like regime, which requires extremely slow rotation of the ambient gas. 
It is unclear whether such conditions are universally realised in the observed systems, whose samples now comprise several tens of objects \cite{Pls2022}. 
In the disc-like regime, a spherically symmetric treatment inevitably overestimates the accretion rate, as the gas becomes concentrated toward the equatorial plane rather than being uniformly distributed in the polar direction.

On the other hand, several subsequent studies \cite{Ranjbar2022, Ranjbar2023} adopted the classical global ADAF formulation  \cite{Narayan1997, Manmoto1997,Yuan2014} to investigate slowly rotating accretion flows. 
However, because this formulation was originally developed for nearly Keplerian accretion discs, it does not recover the Bondi solution in the non-rotating limit.
Notably, a disc-like structure is maintained even when the gas rotation becomes negligible, with a scale height-to-radius ratio (i.e., the disc aspect ratio) of $H/R \sim 0.6$ (see Fig.~1 of \cite{Ranjbar2022}).
As a result, the accretion rate is underestimated, approaching only $\sim 0.54 \dot{M}_{\mathrm{B}}$ in the non-rotating limit rather than converging to $\dot{M}_{\mathrm{B}}$ \cite{Ranjbar2022}.

Observational studies of the X-ray-emitting hot ISM generally indicate that it is quasi-spherical, pressure-supported, and possesses low to moderate angular momentum \cite{Mathews2003, Fabbiano2012, Nardini2022}.
Consequently, ADAFs fed by the hot ISM may span a wide range of configurations, from Bondi-like to disc-like, in realistic astrophysical environments. 
It is therefore essential to establish a unified theoretical framework capable of describing both regimes in a self-consistent manner.

In this paper, we aim to provide such a framework.
By adopting a spherical coordinate system and employing a self-consistent vertical integration scheme, our formulation is designed to smoothly connect the Bondi-like and disc-like limits of accretion.
We demonstrate that our equations recover the Bondi solution in the non-rotating limit and asymptotically approach the classical cylindrical ADAF formulation in the thin-disc limit (section~\ref{consistency}).
We then present a systematic analysis of how key dynamical properties of the flow, including the sonic radius and the mass accretion rate, transition as a function of $\mathcal{L}$, compare our results with previous models, and discuss the observational implications.
\section{Model formulation}\label{model}
\subsection{Basic equations and assumptions}\label{basic_eqs}
We derive our model from the basic equations of hydrodynamics \cite{Mihalas1984}.
We consider a steady ($\partial /\partial t=0$) and axisymmetric ($\partial /\partial \phi=0$) ADAF onto a non-rotating BH at the origin of spherical coordinates $(r,\theta,\phi)$, such that all the physical quantities are functions of $r$ and $\theta$. 
We neglect outflows and set $v_\theta=0$.
The equation of continuity is
\begin{equation}\label{eq_cont}
	\frac{\partial}{\partial r}(r^2\rho v_r)=0,
\end{equation}
where $\rho$ is the mass density and $v_r$ is the radial velocity ($<0$ for inflow).
We assume that the $r\phi$-component of the viscous stress tensor, $t_{r\phi}$, is dominant in the accretion flow (see section~\ref{dis} for further discussion). The three components of the equation of motion can then be written as
\begin{equation}\label{eq_r_motion}
	v_r\frac{\partial v_r}{\partial r} - \frac{v_\phi^2}{r} + \frac{1}{\rho} \frac{\partial p}{\partial r} + \frac{\partial \Psi}{\partial r}=0,
\end{equation}
\begin{equation}\label{eq_t}
	\frac{\partial p}{\partial \theta} - \rho v_\phi^2 \cot\theta=0,
\end{equation}
\begin{equation}\label{eq_phi_motion}
	r^2 \rho v_r \frac{\partial }{\partial r}\left(r v_\phi \right) = \frac{\partial }{\partial r}\left(r^3 t_{r\phi}\right),
\end{equation}
where $v_\phi$ is the azimuthal velocity, $p$ the total pressure, and $\Psi$ the gravitational potential. 
In spherical coordinates, $t_{r\phi}$ is expressed as
\begin{equation}\label{eq_t31}
	t_{r\phi}= \rho \nu r \frac{\partial}{\partial r}\left(\frac{v_\phi}{r}\right).
\end{equation}
where $\nu$ is the kinematic viscosity coefficient. Following previous studies, we adopt the $\alpha$-prescription \cite{SS73}:
\begin{equation}\label{nu}
	\nu=\alpha \frac{c_\mathrm{s}^2}{\Omega_{\mathrm{K}}},
\end{equation}
where $c_\mathrm{s}=\sqrt{p/\rho}$ is the isothermal sound speed.
The effects of general relativity are approximately described with the pseudo-Newtonian potential \cite{PW1980},
\begin{equation}\label{PW_potential}
	\Psi=-\frac{GM}{r-r_\mathrm{g}}.
\end{equation}
The Keplerian velocity and angular velocity are then given by
\begin{equation}
v_\mathrm{K} = \frac{\sqrt{GMr}}{r-r_\mathrm{g}},\qquad
\Omega_{\mathrm{K}} = \frac{v_\mathrm{K}}{r}.
\end{equation}

In the scenarios considered in this paper, the accretion flow is expected to be radiatively inefficient, such that the energy equation can be expressed as
\begin{equation}\label{eq_internalE}
	\rho v_r \left(\frac{\partial e}{\partial r}-\frac{p}{\rho^2} \frac{\partial \rho}{\partial r}\right)= f t_{r\phi} r \frac{\partial}{\partial r}\left(\frac{v_\phi}{r}\right),
\end{equation}
where $e = p/[\rho(\gamma-1)]$ is the specific internal energy of the gas, 
$\gamma$ is the adiabatic index, and $f$ is the advective factor.

Compared with the spherical ADAF model of NF11, our model differs in several key aspects.
First, the $\theta$-component of the momentum equation, our equation~\eqref{eq_t}, is not presented in NF11.
This equation describes the vertical ($\theta$-direction) hydrostatic equilibrium in spherical coordinates and is governed by the balance between the centrifugal force and the pressure gradient in this direction.\footnote{
This differs from cylindrical models, where the vertical hydrostatic equilibrium is governed by the balance between the gravitational force (rather than the centrifugal force) and the pressure gradient in the $z$ direction. 
The reason is that the centrifugal force has no $z$-component, whereas the gravitational force has no $\theta$-component.
}
Since the NF11 model assumed a uniform distribution of physical quantities in the $\theta$ direction, the $\theta$-component of the pressure gradient is zero. Non-zero rotation would then drive the gas towards the equator, until a pressure gradient develops that is sufficient to balance the centrifugal force.
Consequently, high-angular-momentum gas tends to form flattened discs while negligible angular momentum yields nearly spherically symmetric structure, as presented in classical accretion theory \cite{Shap1983,Frank2002,Kato2008}.
While the NF11 model is a good approximation in the case of negligible angular momentum, it is unsuitable in the disc-like case.

Second, while NF11 adopted the remaining hydrodynamic equations, Eqs.~\eqref{eq_cont}, \eqref{eq_r_motion}, \eqref{eq_phi_motion}, and \eqref{eq_internalE}, to construct their model, these equations are formally independent of $\theta$ only when expressed in terms of $v_r$, $c_\mathrm{s}$, and $v_\phi$ (see section~\ref{int_eqs}). 
However, NF11 employed the angular velocity $\Omega$, rather than $v_\phi$, as a primary dynamical variable in their formulation, and further assumed $\theta$-invariant profiles of $v_r$, $c_\mathrm{s}$, and $\Omega$. 
In spherical coordinates, $\Omega = v_\phi /(r \sin\theta)$, so rewriting the equations in terms of $\Omega$ inevitably introduces explicit $\theta$-dependence. 
As a result, a single set of $v_r$, $c_\mathrm{s}$, and $\Omega$ cannot simultaneously satisfy the equations at different $\theta$, unless the rotation is dynamically negligible ($\Omega \to 0$). 
NF11 circumvented this issue by enforcing $\Omega = v_\phi / r$, but the resulting inconsistencies become increasingly significant as rotation plays a more important dynamical role.

We note that NF11 already acknowledged these limitations and indicated that their model is most accurate for $\Omega \ll \Omega_\mathrm{K}$.
Our model aims to establish a unified framework that can describe both Bondi-like and disc-like accretion regimes.
We thus retain the original form of the hydrodynamic equations described above and do not impose spherical symmetry.
This allows for a consistent treatment across a wide range of angular momentum of the ambient gas.
%
\subsection[theta-integrated equations and boundary conditions]{$\theta$-integrated equations and boundary conditions}\label{int_eqs}

To solve the set of equations presented above, a typical treatment is to first obtain (or assume) the vertical distributions of the physical quantities and then integrate the equations in this direction, transforming them into one-dimensional (1D) equations dependent on the radial coordinate.
For example, classical cylindrical ADAF models, formulated in the $(R,\phi,z)$ coordinate system, assume that the radial and azimuthal velocities ($v_R$ and $v_\phi$) are uniform in the $z$ direction, and derive the vertical distributions of $\rho$ and $p$ from the hydrostatic equilibrium equation by adopting one of the one-zone, polytropic, or isothermal assumptions \cite{Narayan1997,Manmoto1997,Gu2007,Xie2008,Gu2009}.

In this subsection, we demonstrate that taking $v_r$, $v_\phi$, and $c_{\mathrm{s}}$ to be $\theta$-independent yields a physically plausible vertical structure. The corresponding $\theta$-profiles of $\rho$ and $p$ follow directly from equation~\eqref{eq_t}. With these profiles, we integrate the governing equations in the $\theta$ direction to obtain a set of 1D radial equations, which can subsequently be solved once appropriate boundary conditions are specified.

From equation~\eqref{eq_cont}, we get
\begin{equation}\label{eq_R1}
\frac{\partial \rho}{\partial r} = -\frac{\rho}{r^2 v_r}\frac{\partial (r^2 v_r)}{\partial r}.
\end{equation}
Substituting equation~\eqref{eq_R1} and $p=\rho c_\mathrm{s}^2$ into Eqs. (\ref{eq_r_motion}), (\ref{eq_phi_motion}), and (\ref{eq_internalE}), we can eliminate $\rho$ from the equations and get
\begin{equation}\label{eq_R2}
v_r\frac{\partial v_r}{\partial r} - \frac{v_\phi^2}{r} + \frac{\partial c_\mathrm{s}^2}{\partial r} -\frac{c_\mathrm{s}^2}{r^2 v_r}\frac{\partial (r^2 v_r)}{\partial r}
+ \frac{GM}{(r-r_\mathrm{g})^2}=0,
\end{equation}
\begin{equation}\label{eq_R3}
r^2 v_r \frac{\partial }{\partial r}\left(r v_\phi \right) = \frac{\partial }{\partial r}\left[\frac{\alpha r^4  c_\mathrm{s}^2}{\Omega_{\mathrm{K}}} \frac{\partial}{\partial r}\left(\frac{v_\phi}{r}\right) \right] - \frac{\alpha c_\mathrm{s}^2 r^2}{\Omega_{\mathrm{K}} v_r} \frac{\partial (r^2 v_r)}{\partial r} \frac{\partial}{\partial r}\left(\frac{v_\phi}{r}\right),
\end{equation}
\begin{equation}\label{eq_R4}
\frac{v_r}{\gamma-1} \frac{\partial c_\mathrm{s}^2}{\partial r} + \frac{c_\mathrm{s}^2}{r^2} \frac{\partial (r^2 v_r)}{\partial r}
= \frac{f \alpha r^2 c_\mathrm{s}^2}{\Omega_{\mathrm{K}}} \left[\frac{\partial}{\partial r}\left(\frac{v_\phi}{r}\right)\right]^2.
\end{equation}
The above three equations govern the dependent variables $v_r$, $c_\mathrm{s}$, and $v_\phi$. Note that these equations are independent of $\theta$ in their mathematical structure. This indicates that if a set of radial profiles of $v_r$, $c_\mathrm{s}$, and $v_\phi$ satisfies the equations on the equatorial plane, then the same set will satisfy the equations at any value of $\theta$.
Therefore, we conclude that a set of $\theta$-independent $v_r$, $c_\mathrm{s}$, and $v_\phi$ satisfies the governing equations of our model.

We note that although a $\theta$-independent $v_\phi$ causes the total centrifugal force to increase as $\theta$ decreases, its radial component remains $v_\phi^2/r$, independent of $\theta$ and equal to the equatorial value. Consequently, the form of the radial momentum equation is preserved at all $\theta$. The $\theta$-component of the centrifugal force is exactly balanced by the $\theta$-direction pressure gradient, ensuring that the momentum equation in the $\theta$-direction is satisfied as well.
When $v_\phi$ is dynamically important, the density profile is naturally concentrated toward the equatorial plane and decreases toward the polar axis (see below); therefore, the polar region contributes negligibly to the $\theta$-integrated equations. On the other hand, when $v_\phi$ becomes small, our setup smoothly converges to a nearly spherical accretion flow, recovering the Bondi-like limit (see section~\ref{limit_Bondi}).

The density and pressure profiles in the $\theta$ direction can now be derived from equation~\eqref{eq_t}, which gives
\begin{equation}\label{rho_t}
	\rho(r,\theta)=\rho_0(r) (\sin \theta) ^{{v_\phi}^2/{c_\mathrm{s}}^2},
\end{equation}
\begin{equation}\label{p_t}
	p(r,\theta)=p_0(r) (\sin \theta) ^{{v_\phi}^2/{c_\mathrm{s}}^2},
\end{equation}
where $\rho_0$ and $p_0$ represent the density and pressure on the equatorial plane and are functions of only $r$. 

We then integrate the governing equations along the $\theta$ direction.
Integrating equation~\eqref{eq_cont} in both the $r$ and $\theta$ directions yields
\begin{equation}\label{eq_mdot}
\begin{split}
    \dot{M} &= -\int_0^\pi 2 \pi r^{2} \rho v_{r} \sin \theta d \theta  \\
    &= -2 \pi r^{2} \rho_0 v_{r} \int_0^\pi (\sin \theta)^{1+\frac{v_\phi^2}{c_\mathrm{s}^2}} d \theta \\
    &= -2 \pi r^{2} \rho_\mathrm{int} v_{r} = \mathrm{const},
\end{split}
\end{equation}
where $\dot{M}$ is taken to be positive for inflow ($v_r<0$), and $\rho_\mathrm{int}$ denotes the density vertically integrated in spherical coordinates, defined as
\begin{equation}\label{rho_int}
\rho_\mathrm{int} \equiv \rho_0 \int_0^\pi (\sin \theta)^{1+\frac{v_\phi^2}{c_\mathrm{s}^2}} d \theta
=\rho_0 \frac{\sqrt{\pi } \Gamma \left(1+\frac{{v_\phi}^2}{2{c_\mathrm{s}}^2}\right)}{\Gamma \left(\frac{3}{2}+\frac{{v_\phi}^2}{2{c_\mathrm{s}}^2}\right)}.
\end{equation}
Here $\Gamma$ represents the $\Gamma$ function.
This expression converges to a spherically symmetric model for $v_\phi \to 0$ and to a disc model when $v_\phi \gg c_{\rm s}$ (see section~\ref{consistency}).

By replacing $p$ with $\rho c_\mathrm{s}^2$ and integrating Eqs. (\ref{eq_r_motion}), (\ref{eq_phi_motion}), and (\ref{eq_internalE}), we derive
\begin{equation}\label{eq_r_motion2}
	v_r\frac{\partial v_r}{\partial r} - \frac{v_\phi^2}{r} + \frac{1}{\rho_\mathrm{int}} \frac{\partial (\rho_\mathrm{int}c_\mathrm{s}^2)}{\partial r} + \frac{\partial \Psi}{\partial r}=0,
\end{equation}
\begin{equation}\label{eq_phi_motion2}
	r^2 \rho_\mathrm{int} v_r \frac{\partial }{\partial r}\left(r v_\phi \right) = \frac{\partial }{\partial r}\left(r^3 T_{r\phi}\right),
\end{equation}
\begin{equation}\label{eq_internalE2}
	\frac{\rho_\mathrm{int} v_r}{\gamma-1} \frac{\partial c_\mathrm{s}^2}{\partial r}- v_r c_\mathrm{s}^2 \frac{\partial \rho_\mathrm{int}}{\partial r}= f T_{r\phi} r \frac{\partial}{\partial r}\left(\frac{v_\phi}{r}\right),
\end{equation}
where
\begin{equation}\label{eq_T31}
	T_{r\phi}= \rho_\mathrm{int} \nu r \frac{\partial}{\partial r}\left(\frac{v_\phi}{r}\right).
\end{equation}
Although these equations closely resemble their two-dimensional counterparts, the dependent variables are now functions of $r$ alone. The problem thus becomes 1D.

In the calculation, it is better to use the differential form of equation~\eqref{eq_mdot}, which is
\begin{equation}\label{eq1I}
	\frac{\partial}{\partial r}(r^2\rho_\mathrm{int} v_r)=0.
\end{equation}
With the help of equation~\eqref{eq1I}, we can integrate equation~\eqref{eq_phi_motion2} in the $r$ direction to get
\begin{equation}\label{eq_j}
	j = r v_\phi - \frac{\alpha r^2 c_\mathrm{s}^2}{v_r \Omega_{\mathrm{K}}} \frac{\partial}{\partial r}\left(\frac{v_\phi}{r}\right)
	=\mathrm{const},
\end{equation}
where $j$ is an integration constant with the dimension of specific angular momentum.
Substituting equation~\eqref{eq1I} into Eqs. (\ref{eq_r_motion2}) and (\ref{eq_internalE2}), we can eliminate $\rho_\mathrm{int}$ and get
\begin{equation}\label{eq_r_motion3}
	v_r\frac{\partial v_r}{\partial r} - \frac{v_\phi^2}{r} + \frac{\partial c_\mathrm{s}^2}{\partial r} -\frac{c_\mathrm{s}^2}{r^2 v_r}\frac{\partial (r^2 v_r)}{\partial r}
	+ \frac{v_\mathrm{K}^2}{r}=0,
\end{equation}
\begin{equation}\label{eq_internalE3}
	\frac{v_r}{\gamma-1} \frac{\partial c_\mathrm{s}^2}{\partial r} + \frac{c_\mathrm{s}^2}{r^2} \frac{\partial (r^2 v_r)}{\partial r}
	= \frac{f \alpha r^2 c_\mathrm{s}^2}{\Omega_{\mathrm{K}}} \left[\frac{\partial}{\partial r}\left(\frac{v_\phi}{r}\right)\right]^2.
\end{equation}
The radial profiles of $v_r$, $c_\mathrm{s}$, and $v_\phi$ can be calculated with Eqs.~\eqref{eq_j}--\eqref{eq_internalE3}. 
The radial profile of $\rho_\mathrm{int}$ can then be calculated using equation~\eqref{eq_mdot}, which scales linearly with the ambient gas density $\rho_\mathrm{out}$. We can actually set $\rho_\mathrm{out}=1$, as it only influences the scaling of $\dot{M}$.
So essentially we have three first-order ODEs with a parameter $j$. 

With Eqs.~\eqref{eq_j}--\eqref{eq_internalE3}, we can calculate the radial derivative of $v_r$, written in a dimensionless form as
\begin{equation}\label{eq_dvrdr}
	\left(\gamma -\frac{{v_r}^2}{{c_\mathrm{s}}^2} \right) \frac{r}{v_r}\frac{\partial v_r }{\partial r} = \frac{v_\mathrm{K}^2 - v_\phi^2 }{c_\mathrm{s}^2} - 2\gamma  
	+ \frac{f (\gamma -1) v_\mathrm{K} v_r (r v_\phi - j)^2}{\alpha r^2 c_\mathrm{s}^4}.
\end{equation}
Because an ADAF is typically subsonic at large radii and must become supersonic near the event horizon, it necessarily undergoes a transonic transition.
Accordingly, the sonic radius $r_\mathrm{c}$ is a critical radius, and 
${\partial v_r}/{\partial r}$ should be well-behaved across $r_\mathrm{c}$. 
This requires that both the numerator and denominator in Eq.~(\ref{eq_dvrdr}) vanish simultaneously at $r=r_\mathrm{c}$:
\begin{equation}\label{D1}
	r=r_\mathrm{c}:\quad \gamma -\frac{{v_r}^2}{{c_\mathrm{s}}^2}=0,
\end{equation}
\begin{equation}\label{N1}
	r=r_\mathrm{c}:\quad \frac{v_\mathrm{K}^2 - v_\phi^2 }{c_\mathrm{s}^2} - 2\gamma  
	+ \frac{f (\gamma -1) v_\mathrm{K} v_r (r v_\phi - j)^2}{\alpha r^2 c_\mathrm{s}^4}=0,
\end{equation}
which provide two boundary conditions at the sonic point.\footnote{It is a common practice in the ADAF literature to refer to the critical point as the "sonic point" \cite{Narayan1997,Manmoto1997,Yuan2014}. In our model, the radial velocity $v_r$ equals the adiabatic sound speed $\sqrt{\gamma} c_\mathrm{s}$ at this point.} 
They also introduce the sonic radius $r_\mathrm{c}$ as an additional parameter to be determined, 
besides the parameter $j$. 
Consequently, three further boundary conditions are still required to close the problem.

We impose two boundary conditions at the outer radius of the accretion flow, assuming that $v_\phi$ and $c_\mathrm{s}$ approach the corresponding values in the surrounding environment.
The last boundary condition is set following NF11: we impose $\partial l/\partial r=0$ ($l=rv_\phi$) at $r_\mathrm{c}$ 
to ensure a smooth transition to the supersonic region. 

With the above boundary conditions, global solutions of our model can be obtained. 
We set $GM=c=1$ and use the relaxation method \cite{Press2002}.

\subsection{Consistency checks and asymptotic limits}\label{consistency}

In this subsection, we demonstrate that our formulation recovers the correct limiting behaviours in both the 
spherical accretion and thin-disc regimes.  
When $v_\phi \rightarrow 0$, our equations reduce to the classical Bondi accretion model.  
In the opposite limit, when $v_\phi \gg c_{\mathrm{s}}$, the accretion flow becomes geometrically thin and our equations 
asymptote to the classical ADAF formulation written in cylindrical coordinates.\footnote{
The classical cylindrical-coordinate ADAF formulation relies on an expansion of the gravitational potential
that is formally valid only in the thin-disc limit 
\cite{Gu2007,Gu2009}.  
Accordingly, full agreement between our formulation and the cylindrical ADAF equations should be expected 
only when the disc is sufficiently thin.  
We note, however, that global ADAF solutions are not strictly thin; their typical aspect ratio is a few tenths.
In this moderately thick regime, numerical comparisons show that the dynamical structure obtained from our 
spherical-coordinate formulation deviates only slightly from that of the classical cylindrical ADAF equations.
}

These consistency checks are not merely formal. 
They ensure that the governing equations remain physically well posed across the entire angular-momentum range, from nearly spherical Bondi accretion to disc-dominated flows, a property that is not generally satisfied by commonly used formulations.  
Together, these checks confirm that our formulation provides a unified and physically consistent description 
bridging the Bondi-like and disc-like limits.

\subsubsection{Spherical limit}\label{limit_Bondi}

When the azimuthal velocity becomes negligible compared with the sound speed, $v_\phi \to 0$,
the centrifugal and viscosity terms disappear from the governing equations.
In this limit, the density and pressure profiles naturally become $\theta$-independent, and the vertical balance equation 
reduces to the trivial identity satisfied by a spherically symmetric flow.
Evaluation of the $\Gamma$ functions (see Appendix~\ref{A1}) in equation~\eqref{rho_int} yields 
$\rho_\mathrm{int}=2\rho_0=2\rho$.
Consequently, equations~\eqref{eq_mdot}, \eqref{eq_r_motion2} and \eqref{eq_internalE2} become
\begin{equation}\label{eq_bondi0}
    \dot{M} = -4 \pi r^{2} \rho v_{r},
\end{equation}
\begin{equation}\label{eq_bondi1}
	v_r\frac{\partial v_r}{\partial r}  + \frac{1}{\rho} \frac{\partial (\rho c_\mathrm{s}^2)}{\partial r} + \frac{\partial \Psi}{\partial r}=0,
\end{equation}
\begin{equation}\label{eq_bondi3}
	\frac{\partial }{\partial r} \left[\frac{c_\mathrm{s}^2}{\rho^{(\gamma-1)}}\right] = 0,
\end{equation}
which are the equations of the classical Bondi model (the last equation is the polytropic relation $p \propto \rho^\gamma$).\footnote{
Classical Bondi solutions often adopt the Newtonian potential, which leads to $r_\mathrm{c}=0$ for $\gamma=5/3$. 
In realistic cases, relativistic effects should be included, and the sonic point lies at a larger radius 
\cite{Michel1972}, typically at hundreds of $r_\mathrm{g}$ in Bondi-like flows
\cite{NF11,Ranjbar2022}.}
Thus, our formulation converges to the Bondi solution in the limit of vanishing rotation.

\subsubsection{Thin-disc limit}

In the opposite regime, where rotation is dynamically dominant, the density becomes strongly concentrated toward the 
equatorial plane, and the $\theta$-profiles of $\rho$ and $p$ described by equations~\eqref{rho_t} and \eqref{p_t} 
become sharply peaked.  
In this limit, our formulation should agree with the classical ADAF equations written in cylindrical coordinates 
\cite{Narayan1997, Manmoto1997, Yuan2014}.

In cylindrical ADAF models, the disc aspect ratio is characterised by 
$H/r = c_{\mathrm{s}}/v_{\mathrm{K}}$,\footnote{
In the thin-disc regime, the flow is concentrated near the equatorial plane, 
where the spherical radial coordinate $r$ and the cylindrical radial coordinate $R$ 
satisfy $R = r \sin\theta \simeq r$.  
We therefore use $r$ for convenience throughout this subsection.
}
and is not directly expressed in terms of $v_\phi$. 
Since rotation typically approaches Keplerian in the limit of a geometrically thin flow,
here for simplicity we parameterise the rotation as $v_\phi=\xi v_{\mathrm{K}}$ with $\xi\sim 1$.

Under this condition ($v_\phi = \xi v_{\mathrm{K}} \gg c_{\mathrm{s}}$),
evaluation of the $\Gamma$ functions (see Appendix~\ref{A2}) in equation~\eqref{rho_int} yields
\begin{equation}\label{rho_int_thin}
\rho_\mathrm{int} 
=  \frac{\rho_0 \sqrt{\pi}}{\sqrt{\frac{{v_\phi}^2}{2{c_\mathrm{s}}^2}}}
= \rho_0 \frac{\sqrt{2 \pi}}{\xi} \frac{H}{r}.
\end{equation}
Consequently, equation~\eqref{eq_mdot} becomes
\begin{equation}\label{eq_mdot_thin}
    \dot{M} 
    = -4 \pi r H \rho_0 v_{r}\cdot \frac{1}{\xi}\sqrt{\frac{\pi}{2}},
\end{equation}
which is the familiar form of the mass accretion rate in cylindrical models, with 
$\sqrt{\pi/2}/\xi$ being a correction factor from vertical integration. 
For $\xi=1$, this factor reduces to that obtained by Manmoto et al. \cite{Manmoto1997}.

The radial momentum equation and the energy equation, Eqs.~\eqref{eq_r_motion} and \eqref{eq_internalE}, are already identical to those in cylindrical formulations, even before vertical integration.
The $\theta$-integrated angular momentum equation, Eq.~\eqref{eq_phi_motion2}, becomes
\begin{equation}\label{eq_phi_thin}
	r H \rho_0 v_r \frac{\partial }{\partial r}\left(\Omega r^2 \right) 
	= \frac{\partial }{\partial r}\left(r^3 H \nu \rho_0 \frac{\partial \Omega}{\partial r} \right),
\end{equation}
where $\Omega=v_\phi/r$ in the thin-disc limit.  
This is precisely the standard angular-momentum equation used in the cylindrical ADAF models.

We thus conclude that, in the thin-disc limit, our formulation asymptotically agrees with the classical cylindrical-coordinate ADAF equations.
%
\section{Numerical results}\label{results}
\subsection{Accretion regimes as a function of angular momentum}

We solve our model numerically using the relaxation method.
For comparison, we also solve the spherically symmetric model of NF11 and the classical ADAF model, adopting identical input parameters and ambient gas properties.
The outer boundary is set at the Bondi radius,
\begin{equation}\label{rB}
    r_\mathrm{B} \equiv \frac{GM}{c_\mathrm{out}^2},
\end{equation}
where $c_\mathrm{out}$ is the sound speed of the ambient gas.
We adopt $\alpha=0.1$, $f=1$, $\gamma=5/3$, and $c_\mathrm{out}=10^{-3}c$,
corresponding to an ambient gas temperature
$T_\mathrm{out}=6.5\times10^{6}$ K.
These values are commonly adopted in previous studies
\cite{NF11,Ranjbar2022}
and are appropriate for hot ISM accretion onto SMBHs
at the centres of elliptical galaxies.
The remaining outer boundary condition is the azimuthal velocity
$v_\phi$ at $r_\mathrm{out}=r_\mathrm{B}$, which is specified through
the dimensionless angular momentum parameter $\mathcal{L}$ as
\begin{equation}\label{vphi_out}
    v_\phi(r_\mathrm{B})=\frac{\mathcal{L}  l_\mathrm{ms}}{r_\mathrm{B}}.
\end{equation}

\begin{figure*}
\centering
\includegraphics[width=\textwidth]{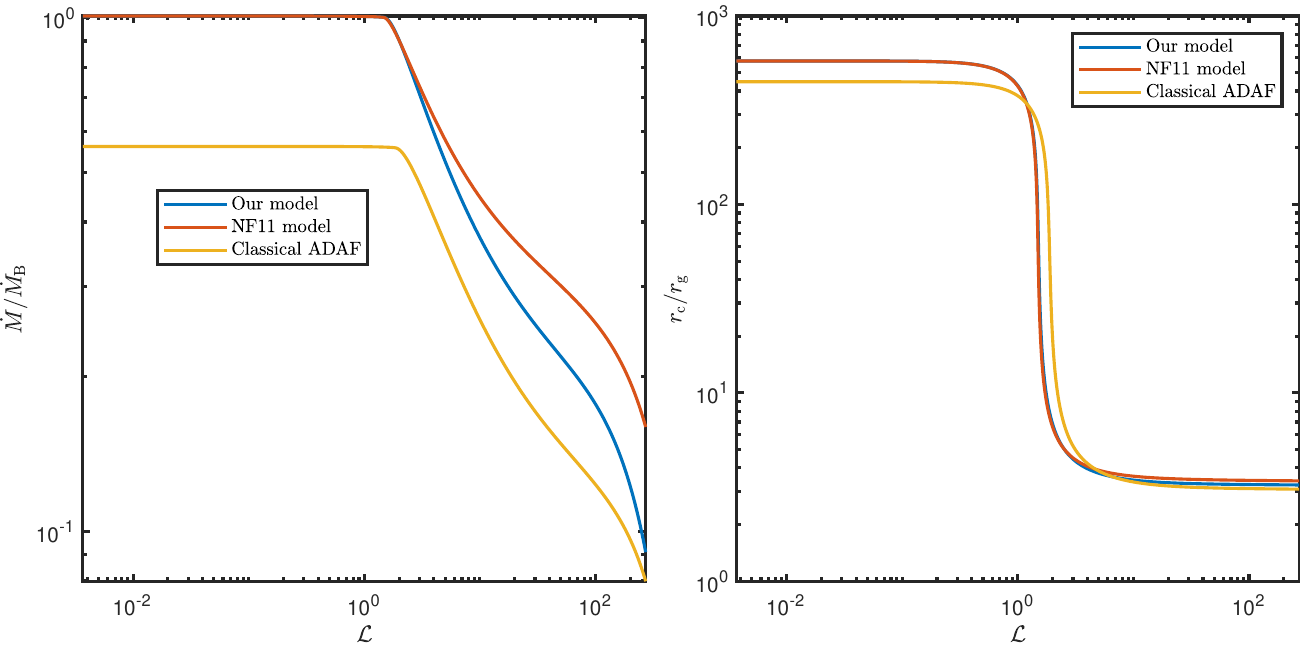}
\caption{Accretion rate $\dot{M}$ (left panel) and sonic radius $r_\mathrm{c}$ (right panel) as functions of $\mathcal{L}$, for $\alpha=0.1$, $f=1$, $\gamma=5/3$, and $c_{\mathrm{out}}=10^{-3}c$.
The blue, red, and yellow lines correspond to our model, the NF11 model, and the classical ADAF model, respectively.
The horizontal axis spans from $\mathcal{L} \ll 1$ to $\mathcal{L} = 272$, the latter corresponding to Keplerian rotation at $r_\mathrm{B}$.
}
\label{fig1}
\end{figure*}

Our primary goal here is to demonstrate that our model provides a unified description of both Bondi-like and disc-like accretion regimes.
This can be clearly illustrated by Fig.~\ref{fig1},
which shows the dependence of the accretion rate $\dot{M}$
and the sonic radius $r_\mathrm{c}$ on $\mathcal{L}$.\footnote{Each point on the curves corresponds to a global solution; the curves appear smooth because we computed thousands of global solutions for each model.
}
The blue, red, and yellow lines correspond to our model, the NF11 model, and the classical ADAF model, respectively.
The parameter $\mathcal{L}$ spans from $\mathcal{L} \ll 1$ to $\mathcal{L} = 272$, the latter corresponding to Keplerian rotation at $r_\mathrm{B}$.
The accretion rate is normalised by the Bondi accretion rate \cite{Frank2002},
\begin{equation}\label{Bondi_Mdot}
    \dot{M}_\mathrm{B}=\pi G^2 M^2 \frac{\rho_\mathrm{out}}
    {\gamma^{3/2}c_{\mathrm{out}}^3}
    \left(\frac{2}{5-3 \gamma}\right)^{\frac{5-3 \gamma}{2(\gamma-1)}}.
\end{equation}
For $\gamma = 5/3$, the $\gamma$-dependent exponential factor equals unity, yielding
$\dot{M}_\mathrm{B} = \pi G^2 M^2 \rho_\mathrm{out} / (\gamma^{3/2} c_\mathrm{out}^3)$.
Note that the dependence on $\rho_\mathrm{out}$ cancels out
when considering the ratio $\dot{M}/\dot{M}_\mathrm{B}$.

Figure~\ref{fig1} shows that all three models exhibit two distinct
regimes in their solutions as the dimensionless angular momentum
parameter $\mathcal{L}$ varies. The transition occurs roughly around 
$\mathcal{L} \sim 1.5$--$2$ in the NF11 and our models, consistent with the range 
identified by NF11, while in the classical ADAF model the transition occurs at slightly higher $\mathcal{L}$.
Despite this qualitative similarity, the physical interpretation of
these regimes and the predicted accretion rates differ substantially
among the models.

In the NF11 model, the high-accretion-rate branch at low $\mathcal{L}$ converges to the Bondi solution: both the mass accretion rate and the sonic radius approach their Bondi values.\footnote{ Compared with the accretion rate, the sonic radius $r_\mathrm{c}$ in Bondi-like flows is more sensitive to the choice of the outer boundary $r_\mathrm{out}$. In the limit $\mathcal{L}\to 0$, the values of $r_\mathrm{c}$ obtained from the Bondi model, our model, and the NF11 model converge to the same value when computed with the same outer boundary $r_\mathrm{out}=r_\mathrm{B}$. This value differs slightly from that reported in NF11, who adopted a larger outer radius ($r_\mathrm{out} \lesssim 10 r_\mathrm{B}$).} For this reason, NF11 referred to this branch as Bondi-like. As $\mathcal{L}$ increases, the sonic radius moves inward and the flow properties tend toward those characteristic of an accretion disc, suggesting a transition to a disc-like regime. However, because the NF11 model enforces spherical symmetry, it cannot self-consistently describe the intrinsically non-spherical flow structure in this high-$\mathcal{L}$ regime, leading to an overestimation of the accretion rate.

On the other hand, the classical ADAF model, based on the thin-disc approximation \cite{Gu2007,Gu2009}, yields a disc-like flow structure over the entire range of $\mathcal{L}$. 
Although a high-accretion-rate regime also exists at low $\mathcal{L}$, neither the accretion rate nor the sonic radius fully recovers the Bondi values in the non-rotating limit $\mathcal{L} \to 0$. 
Previous work adopting this model \cite{Ranjbar2022} referred to this regime as Bondi-like; however, the flow actually remains a slim, disc-like structure, with $H/R \sim 0.6$.

In contrast, our model converges to the Bondi solution in the low-$\mathcal{L}$ limit and approaches the classical ADAF solution as $\mathcal{L}$ increases. It therefore captures both Bondi-like and disc-like accretion behaviour within a single, self-consistent framework.
%
\subsection{Dependence on the viscosity parameter and observational implications}

Assuming that a fraction of the accretion power, at the level of up to $\sim 5\%$, can be converted into jet mechanical energy, the observed jet power of $\sim 0.02\dot{M}_\mathrm{B} c^2$ \cite{Allen2006} suggests that the mass accretion rate onto the SMBH should satisfy $\dot{M} \gtrsim 0.4 \dot{M}_\mathrm{B}$. For the solutions obtained with our model and shown in Fig.~\ref{fig1}, this requirement corresponds to $\mathcal{L} \lesssim 8.5$ (or $\mathcal{R} \lesssim 0.03$), implying that the angular velocity of the ambient gas must be limited to only a few per cent of the Keplerian value at the Bondi radius. This places a rather stringent constraint on the rotation of the hot ISM.

\begin{figure*}
\centering
\includegraphics[width=\textwidth]{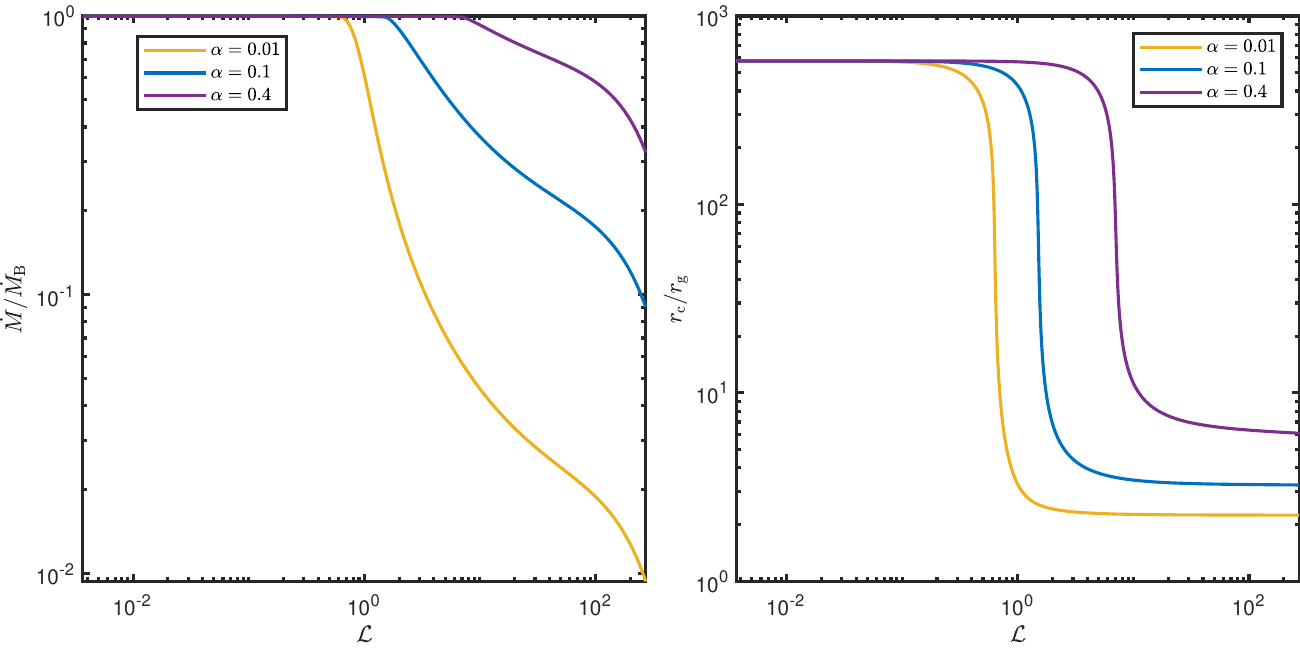}
\caption{Accretion rate $\dot{M}$ (left panel) and sonic radius $r_\mathrm{c}$ (right panel) as functions of $\mathcal{L}$ for different values of $\alpha$ in our model: $\alpha=0.01$ (yellow), 0.1 (blue), and 0.4 (purple).
Other parameters are $f=1$, $\gamma=5/3$, and $c_{\mathrm{out}}=10^{-3}c$.}
\label{fig2}
\end{figure*}

The above restriction, however, depends sensitively on the viscosity parameter $\alpha$.
Figure~\ref{fig2} illustrates how the accretion solutions vary with $\mathcal{L}$ for different values of $\alpha$.
As $\alpha$ increases, the solutions remain Bondi-like up to larger values of $\mathcal{L}$, and the accretion rate in the disc-like regime also increases for a fixed $\mathcal{L}$.
Observationally, values of $\alpha \sim 0.1$--$0.4$ have been proposed \cite{King2007}.
In particular, for $\alpha = 0.4$, the condition $\dot{M}/\dot{M}_\mathrm{B} \gtrsim 0.4$ is satisfied for $\mathcal{L} \lesssim 216$ (or $\mathcal{R} \lesssim 0.79$), corresponding to an ambient gas rotation that is comparable to the Keplerian value at $r_\mathrm{B}$.
This represents a much less restrictive requirement and suggests that, for sufficiently efficient angular momentum transport (as quantified by a large $\alpha$), accretion rates at a significant fraction of the Bondi rate can be sustained even in the presence of substantial ambient gas rotation.

Note that if the jet production efficiency were higher than 5\%, the required accretion rate could be correspondingly lower, which would further relax the constraints on the ambient gas rotation. 

\section{Discussion}\label{dis}
\subsection{On the viscous stress tensor}

In this work, we consider only the $r\phi$-component of the viscous stress tensor, $t_{r\phi}$, and neglect the other components.
In our formulation, the angular velocity is not uniform in the $\theta$ direction, which in principle allows for a non-zero $\theta\phi$-component, $t_{\theta\phi}$.
Nevertheless, we argue that neglecting $t_{\theta\phi}$ remains a reasonable approximation for the following reasons.

First, owing to reflection symmetry about the equatorial plane, $t_{\theta\phi}$ vanishes at the midplane and becomes significant only at high latitudes.
In the disc-like regime, the accreted material is strongly concentrated toward the equatorial plane, so the contribution from high-latitude regions is subdominant, making the neglect of $t_{\theta\phi}$ a reasonable approximation.
In the Bondi-like regime, although material at high latitudes becomes dynamically important, the rotation itself is negligible and the flow becomes nearly $\theta$-independent, which again renders $t_{\theta\phi}$ insignificant.

Second, viscous stresses in accretion flows are now widely believed to arise from magneto-rotational instability (MRI; \cite{Zhu2018,Oliver2019,Oliver2022,Jiang2023}).
Using global magnetohydrodynamical simulations, Zhu \& Stone \cite{Zhu2018} systematically investigated the relative importance of different components of the viscous stress tensor and found that approximately $95\%$ of the disc accretion is due to the $r$--$\phi$ stress.
This result lends strong support to our simplified treatment.

We acknowledge that neglecting $t_{\theta\phi}$ remains an approximation, particularly in the disc-like regime, and that including this component may modify the detailed vertical structure of the flow.
However, it is not expected to alter the global accretion properties determined by the vertically integrated equations.
A more complete investigation of the full viscous tensor will be pursued in future work.

We also comment on the $rr$-component of the viscous stress tensor, $t_{rr}$.
As discussed by NF11, the classical Bondi model---serving as the natural reference for slowly rotating accretion flows---neglects $t_{rr}$.
Following this reasoning, and in order to facilitate direct comparison with both Bondi accretion and previous studies, we likewise neglect $t_{rr}$ in the present work.
The effect of this term in both Bondi solutions and slowly rotating accretion flows is left for future investigation.
%
\subsection{Role of the galactic gravitational potential}

An implicit assumption in our calculations, following NF11, is that the gravitational potential is dominated by the SMBH within the radial range of interest, and that the contribution from the host galaxy can be neglected. This assumption is expected to hold near the Bondi radius, where the SMBH gravity sets the characteristic scale of the accretion flow.

As discussed by NF11, at radii significantly larger than the Bondi radius the gravitational potential of the host galaxy may become important or even dominant \cite{QN2000}. To avoid explicitly modelling this regime, NF11 restricted the outer boundary to be no more than a factor of a few larger than the Bondi radius, typically $r_\mathrm{out} \lesssim 10 r_\mathrm{B}$, within which the galactic potential can be safely neglected.

In this work, we adopt the same strategy and focus on the accretion flow within the Bondi radius, setting $r_\mathrm{out}=r_\mathrm{B}$. Within this region, the omission of the galactic gravitational potential is not expected to qualitatively alter the accretion rate or the global flow properties. Indeed, recent calculations that explicitly include the galactic gravitational potential within the classical ADAF framework \cite{Ranjbar2022} show that the resulting changes in the accretion rate near the Bondi radius are typically at the level of $\sim 10\%$ for low angular momentum flows.

We therefore conclude that neglecting the galactic gravitational potential is a reasonable approximation for the purpose of this study, which aims to understand how angular momentum regulates accretion near the Bondi radius and to assess whether accretion rates at a significant fraction of the Bondi rate can be maintained under realistic conditions.

%
\subsection{Outflows and observational constraints}

Another important uncertainty in hot accretion flows concerns the possible presence of strong outflows. Early analytical studies of ADAFs noted that the Bernoulli parameter can be positive, suggesting that the gas may be gravitationally unbound and capable of driving powerful winds \cite{NY94,Narayan1995a}. On this basis, Blandford \& Begelman \cite{BB99} proposed that the mass accretion rate could decrease sharply with decreasing radius, potentially reducing the accretion rate at the black hole by orders of magnitude. Subsequent numerical simulations \cite{Igu2000, Yuan2012, Li2013, Yuan2015} have indeed found that the mass inflow rate decreases inward, although there remains no consensus on whether the gas truly escapes to infinity or on the physical interpretation of the Bernoulli parameter \cite{Abr2000,Yuan2014}.

In the context of X-ray luminous elliptical galaxies, however, strong outflows pose a serious challenge when confronted with observations.
As emphasized by NF11, the tight correlation between the Bondi accretion rate and the observed jet power \cite{Allen2006} implies that the actual accretion rate onto the SMBH must be comparable to, or at least a substantial fraction of, the Bondi rate, adopting a jet production efficiency of a few per cent.
NF11 further examined the energetic implications of different jet launching mechanisms and concluded that, if mass loss through outflows were as severe as suggested in some phenomenological models, there would simply be insufficient accretion power available to drive the observed jets.
From this perspective, the key difficulty is not an excess of accretion,
but rather maintaining a sufficiently high accretion rate.

For this reason, following NF11, we neglect mass loss through outflows in the present work. This assumption is not meant to imply that outflows are entirely absent, but rather that they must be dynamically subdominant in the systems of interest. We note that phenomenological prescriptions in which the accretion rate scales as a power law with radius have been applied to classical ADAF models even in the low-angular-momentum regime \cite{Ranjbar2023}.
However, the physical consistency of such prescriptions for nearly spherical, Bondi-like flows remains unclear, since the geometry and launching mechanism of outflows in this limit are not well understood. Moreover, strong mass loss in these models typically leads to accretion rates far below the Bondi value, in tension with the observational requirements discussed above.

We therefore conclude that, for the purpose of explaining the observed
jet powers in X-ray luminous elliptical galaxies, it is both reasonable
and necessary to focus on solutions in which outflows are weak and the
accretion rate remains at least several tenths of the Bondi prediction.
%
\section{Summary}\label{sum}

In this paper, we develop a unified framework for hot, radiatively inefficient accretion flows that consistently describes both Bondi-like and disc-like regimes within a single, self-consistent model.
This study is motivated by observations of X-ray luminous elliptical galaxies showing a close correspondence between the Bondi accretion rate and the mechanical power of relativistic jets, as well as by the long-standing tension between classical Bondi accretion, applicable to nearly spherical flows, and standard ADAF models, which describe rapidly rotating accretion discs but become ill-defined in the Bondi-like regime.

Our framework smoothly connects these two classical limits.
In the regime of very low angular momentum, the solutions naturally reduce to the Bondi accretion flow, recovering both its characteristic accretion rate and near-spherical geometry.
At high angular momentum, the solutions converge to the properties of classical ADAFs, including rotational support and a disc-like structure, demonstrating consistency with the standard picture of hot accretion in rapidly rotating systems.

Crucially, our model provides a continuous and physically well-defined description in the intermediate-angular-momentum regime, where neither the Bondi solution nor classical ADAF formulations are strictly applicable.
In this sense, the framework resolves the complementary limitations of previous approaches: it remains valid at high angular momentum, where the spherical symmetry assumed in the NF11 model breaks down, and at low angular momentum, where the disc-like structure predicted by classical ADAF models becomes unphysical.

We further show that the accretion rate is strongly regulated by both the angular momentum of the ambient gas and the viscosity parameter $\alpha$, which characterizes the efficiency of angular momentum transport.
For sufficiently large (yet still reasonable) values of $\alpha$, the accretion rate can remain a substantial fraction of the Bondi rate even in the presence of significant ambient gas rotation, thereby accounting for the observed jet powers.
This provides a natural physical interpretation of how accretion rates comparable to, or at least a substantial fraction of, the Bondi rate can be sustained in realistic galactic environments with finite angular momentum.

Overall, our results establish a unified and flexible framework for hot accretion flows that bridges classical Bondi accretion and ADAF theory.
They offer a coherent physical basis for interpreting accretion and feedback in X-ray luminous elliptical galaxies.
While motivated by these systems, the framework developed here is generic and can be applied to any hot, pressure-supported accretion environment with finite angular momentum.
It also provides a foundation for future extensions incorporating additional physics such as outflows, magnetic fields, and galactic gravitational potentials.
\section*{Acknowledgements}

This work was supported by the Science Foundation of Yunnan Province
(Nos. 202401AS070046 and 202503AP140013), the International Partnership
Program of the Chinese Academy of Sciences (No. 020GJHZ2023030GC),
and the Yunnan Revitalization Talent Support Program.


\bibliographystyle{unsrt}   
\bibliography{refs}
\appendix

\section{Asymptotic Expansions of the Ratio of Gamma Functions}
\label{app_gamma}

In this appendix we derive the limiting behaviour of the ratio
\begin{equation}
R(x) = \frac{\Gamma(1+x)}{\Gamma\left(\frac{3}{2}+x\right)},
\end{equation}
in the limits $x \to 0$ and $x \to \infty$ for $x>0$.

\subsection[Limit as x approaches 0]{Limit as $x \to 0$}
\label{A1}

Since the Gamma function is analytic for $\mathrm{Re}(z)>0$, the limit
as $x \to 0$ can be evaluated by direct substitution:
\begin{equation}
\lim_{x \to 0} R(x)
= \frac{\Gamma(1)}{\Gamma\left(\frac{3}{2}\right)}
= \frac{1}{\frac{1}{2}\sqrt{\pi}}
= \frac{2}{\sqrt{\pi}}.
\end{equation}

\subsection[Limit as x approaches infinity]{Limit as $x \to \infty$}
\label{A2}

The asymptotic behaviour of the ratio of Gamma functions is given by
the well-known expansion
\begin{equation}
\label{eq_St}
\frac{\Gamma(x+a)}{\Gamma(x+b)}
\sim x^{a-b}
\left[1+\mathcal{O}\left(\frac{1}{x}\right)\right],
\quad \text{as } x \to \infty,
\end{equation}
which follows directly from Stirling’s approximation.

Applying equation~(\ref{eq_St}) with $a=1$ and $b=3/2$ yields
\begin{equation}
R(x) = \frac{\Gamma(1+x)}{\Gamma\left(\frac{3}{2}+x\right)}
\sim x^{-\frac{1}{2}}
\left[1+\mathcal{O}\left(\frac{1}{x}\right)\right],
\quad \text{as } x \to \infty.
\end{equation}

\end{document}